\documentclass[conference,letterpaper]{IEEEtran}

\usepackage[utf8]{inputenc} 
\usepackage[T1]{fontenc}
\usepackage{url}
\usepackage{ifthen}
\usepackage{cite}
\usepackage[cmex10]{amsmath} %
\usepackage{amssymb} 
\usepackage{mathrsfs}
\usepackage[mathcal]{euscript}

\usepackage{tikz}
\usetikzlibrary{shapes,arrows}
\usetikzlibrary{calc}
\allowdisplaybreaks

\newcommand{\secref}[1]{Section~\ref{#1}}

\newcommand{\figref}[1]{Fig.~\ref{#1}}
\newcommand{\lemref}[1]{Lemma~\ref{#1}}
\newcommand{\thmref}[1]{Theorem~\ref{#1}}
\newcommand{\propref}[1]{Proposition~\ref{#1}}

\newcommand{\cgk}{C_{\mathrm{GK}}}
\newcommand{\cw}{C} %
\newcommand{\cdk}{\mathfrak{i}} %

\newcommand{\cU}{\mathcal{U}}
\newcommand{\cS}{\mathcal{S}}
\newcommand{\cT}{\mathcal{T}}
\newcommand{\cW}{\mathcal{W}}
\newcommand{\cX}{\mathcal{X}}
\newcommand{\cY}{\mathcal{Y}}
\newcommand{\cZ}{\mathcal{Z}}

\newcommand{\Wb}{\bar{W}}

\newcommand{\fh}{\hat{f}}
\newcommand{\gh}{\hat{g}}

\newcommand{\ev}{{\mathbb{E}}}
\newcommand{\pr}{\mathbb{P}}
\newcommand{\prob}[1]{\pr\left\{#1\right\}}

\newcommand{\E}[1]{\ev\left[#1\right]}

\newcommand{\T}{\mathrm{T}} %

\newcommand{\fss}{f^*}
\newcommand{\gss}{g^*}

\newcommand{\mcf}[1]{f^*_{#1}}
\newcommand{\mcg}[1]{g^*_{#1}}

\newcommand{\defeq}{\triangleq}
\newcommand{\ds}{\displaystyle}

\newtheorem{definition}{Definition}
\newtheorem{corollary}{Corollary}
\newtheorem{theorem}{Theorem}
\newtheorem{proposition}{Proposition}
\newtheorem{lemma}{Lemma}
\newtheorem{remark}{Remark}

\newcommand{\ibls}{L^*_{\mathrm{IB}}}

\begin{document}
\title{Separable Computation of Information Measures}

\author{%
  \IEEEauthorblockN{Xiangxiang Xu and Lizhong Zheng}
  \IEEEauthorblockA{
    Dept. EECS, MIT\\
    Cambridge, MA 02139, USA\\
    Email: \{xuxx, lizhong\}@mit.edu}
}%

\maketitle

\begin{abstract}
   We study a separable design for computing information measures, where the information measure is computed from learned feature representations instead of raw data. Under mild assumptions on the feature representations, we demonstrate that a class of information measures admit such separable computation, including mutual information, $f$-information, Wyner's common information, G{\'a}cs--K{\"o}rner common information, and Tishby's information bottleneck. Our development establishes several new connections between information measures and the statistical dependence structure. The characterizations also provide theoretical guarantees of practical designs for estimating information measures through representation learning.
\end{abstract}

\tikzset{%
  block/.style    = {draw, thick, rectangle, minimum height = 3em, minimum width = 5em},
  extfill/.style={fill = blue!20}, %
  frozenfill/.style={fill = gray, fill opacity = .4}, %
  ext/.style={trapezium, trapezium angle=67.5, draw,
  inner ysep=5pt, outer sep=0pt, extfill,
  minimum height=1.2cm, minimum width=0pt},
  input/.style    = {coordinate}, %
}

\section{Introduction}
The computation of information measures, e.g., mutual information, is a fundamental task in information theory  \cite{shannon1951prediction} and its applications. In machine learning tasks, for example, computing information measures becomes a key step when using information-theoretic tools for algorithm analyses and designs. 
Due to the unknown statistical model behind practical data, information measures cannot be directly computed from the definitions. The high dimensionality and complicated structures of data also lead to an enormous computation complexity of non-parametric approaches. %

Recent developments employed deep learning techniques in designing information measure estimators \cite{belghazi2018mutual, gowri2024approximating}, leveraging the capability of deep neural networks in effectively processing high-dimensional and structured data. For instance, \cite{gowri2024approximating} proposed an approach to estimate mutual information by using deep neural networks to extract features from data and then applying a non-parametric mutual information estimator on the extracted features. Despite their performance gains on empirical evaluations, such designs are often heuristic without guarantees, where the learned features may fail to carry useful information for the estimation tasks \cite[Sec. 3.3]{gowri2024approximating}. %

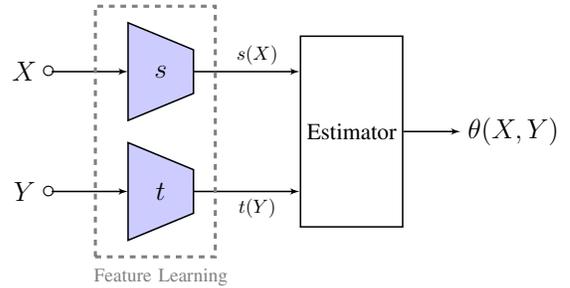
\begin{figure}[!t]
  \centering
  \resizebox{.85\columnwidth}{!}{\begin{tikzpicture}[auto, thick, node distance=2cm, >=latex'
  ]
  \def\dy{1.1}

  \draw node [block, text centered, minimum height = 3.5cm] (E) at (3.5, 0)  {\large Estimator};
  \draw [->] (E) -- + (2, 0) node [right] {\Large $\theta(X, Y)$};

  \foreach \i/\ti/\tu/\txtA/\loc [count = \c]  in {1/$X$/$S$/$s(X)$/above, {-1}/$Y$/$T$/$t(Y)$/below} 
  {
    \draw node [ext, rotate = -90]  (f-\c) at (0, \dy * \i cm){};
    \draw node [input, left of = f-\c] (x-\c) {}
    node [xshift = -0.5cm] at (x-\c) {\Large \ti}
    node [xshift = -0.6mm] at (x-\c) {\Large $\circ$};
    \draw [->] (x-\c) -- (f-\c);
    \draw [->] (f-\c.north) -- (E.west|-f-\c.north) node [\loc, pos = .6] { \txtA};
  }

  \draw node at (f-1)  {\Large $s$}
  node at (f-2)  {\Large $t$};

  \draw [draw=gray,opacity = .8, dashed, ultra thick] ($(f-1) + (-1.2, 1.2)$) rectangle ($(f-2)+(1, -1.2)$) {} node [gray, below of = f-2, yshift = 1.2em] {Feature Learning};

\end{tikzpicture}

}
  \caption{Computing an information measure $\theta(X, Y)$ in two steps: (1) obtain transformed variables $s(X), t(Y)$ and (2) apply an estimator on $s(X), t(Y)$. For high-dimensional $X$, $Y$ with unknown probability structures, the first step can be implemented by data-driven approaches, e.g., deep neural network training, corresponding to the feature learning process, where $s(X)$ and $t(Y)$ are the learned feature representations.}
  \label{fig:blk}
\end{figure}

In this paper, we formulate the problem of computing information measures from extracted features, which allows separable computation and implementation. We restrict to the bivariate case and consider computing an information measure $\theta(X, Y)$ from a pair of random variables $X$ and $Y$, where $X$ or $Y$ correspond to possibly high-dimensional data in practical applications. As illustrated in \figref{fig:blk}, the separable computation is conducted by first learning features $s(X)$, $t(Y)$ from $X, Y$, and then applying $\theta$-estimator on the extracted features. Our focus is on establishing the theoretical conditions such that the features contain the necessary information for estimating the $\theta$, i.e., $\theta(X, Y) = \theta(s(X), t(Y))$. 
In particular, we assume that $s(X)$ and $t(Y)$ can be any sufficient statistics of $X$ and $Y$, i.e., we have the Markov relations $X-s(X)-Y$ and $X-t(Y)-Y$. %
This assumption allows flexible feature learning designs. Moreover, the statistical sufficiency allows feature extractors to discard irrelevant information while capturing the statistical dependence between $X$ and $Y$ \cite{xu2024dependence}. %
Under this assumption, we establish several classical information measures admitting the separable computation, including mutual information and its variants using $f$-divergence, Wyner's common information \cite{wyner1975common}, G{\'a}cs--K{\"o}rner common information \cite{gacs1973common}, and Tishby's information bottleneck \cite{tishby2000information}. Our results provide theoretical guarantees of modular designs in applying deep neural networks for information measure estimations, which can lead to more systematic designs and improve
implementation efficiency. %

\section{Preliminaries and Notations}
For a random variable $Z$, we denote the corresponding alphabet by $\cZ$, use $z$ to denote a specific value in $\cZ$, and use $P_Z$ to denote the probability distribution. Throughout our development, we restrict to a pair of discrete random variables $X$, $Y$ on finite alphabets $\cX, \cY$. Without loss of generality, we assume $P_X$ and $P_Y$ have positive probability masses, i.e., $P_{X}(x) > 0$ for all $x \in \cX$ and $P_Y(y) > 0$ for all $y\in \cY$, since otherwise we can remove the zero-probability symbols from the alphabets. For $n \geq 1$, we denote $[n] \defeq \{1, 2, \dots, n\}$.

\subsection{Canonical Dependence Kernel and Modal Decomposition}%
Given $(X, Y) \sim P_{X, Y}$ the associated \emph{canonical dependence kernel} (CDK) function \cite{HuangMWZ2024} is defined as a joint function $\cdk_{X; Y}$:
\begin{align}
  \cdk_{X; Y}(x, y) \defeq \frac{P_{X, Y}(x, y)}{P_{X}(y)P_Y(y)} - 1.
  \label{eq:cdk:def}
\end{align}

By applying the singular value decomposition (SVD) in the space of joint function, we can write CDK $\cdk_{X; Y}$ as a superposition of rank-one singular modes, 
referred to as its modal decomposition \cite{xu2024neural}:
\begin{align}%
  \cdk_{X; Y}(x, y) = \sum_{i = 1}^K  \sigma_i \cdot \mcf{i}(x) \mcg{i}(y),
  \label{eq:md:def}
\end{align}
 In particular, $\sigma_1 \geq \sigma_2 \geq \dots \geq \sigma_K > 0$ are the singular values, where $K \geq 0$ denotes the rank of $\cdk_{X; Y}$. Analogous to the orthogonality of singular vectors in matrix SVD, $\mcf{i}, \mcg{i}$ satisfy $\E{\mcf{i}(X) \mcf{j} (X)} = \E{\mcg{i}(Y) \mcg{j}(Y)} = \delta_{i, j}$, for all $ i, j \in [K]$. %
 For convenience, we also define %
 functions
 \begin{align}
   \fss \defeq (\mcf{1}, \dots, \mcf{K})^\T, \quad \gss \defeq (\mcg{1}, \dots, \mcg{K})^\T.\label{eq:fss:gss}  
 \end{align}%

For convenience, we further define $\sigma_0 \defeq 1$, $\mcf{0} \colon \cX \ni x \mapsto 1$ and $\mcg{0}\colon \cY \ni y \mapsto 1$. Then, we can write the SVD of the density ratio as [cf. \eqref{eq:cdk:def}]:
 $\ds \frac{P_{X, Y}(x, y)}{P_X(x)P_Y(y)} = \sum_{i = 0}^K  \sigma_i \cdot \mcf{i}(x) \mcg{i}(y).$

\subsection{Sufficiency and Minimal Sufficiency}
\begin{definition}
 Given $X, Y$, $S = s(X)$ is a sufficient statistic of $X$ for inferring $Y$ if we have the Markov relation $X-s(X)-Y$. We call $f(X)$ a minimal sufficient statistic if $f(X)$ is sufficient, and for each given sufficient statistic $S$ of $X$, $f(X)$ is a function of $S$.
\end{definition}

We have the following useful characterization of sufficient statistics. A proof is provided in \secref{proof:prop:suff}.%

\begin{proposition}
  \label{prop:suff}
  Given $X, Y$ and $S = s(X), T = t(Y)$, the following statements are equivalent:
\begin{itemize}
\item $X-S-Y$, $X-T-Y$;
\item $X-S-T-Y$;
\item $\ds \frac{P_{X, Y}(x, y)}{P_X(x)P_Y(y)} = \frac{P_{S, T}(s(x), t(y))}{P_S(s(x))P_T(t(y))}$ for all $x \in \cX, y \in \cY$.
\end{itemize}
\end{proposition}
From such equivalences, the Markov relation $X-S-T-Y$ is equivalent to $X-S-Y$ and $X-T-Y$, i.e., both $S$ and $T$ are sufficient statistics.

In addition, the following result demonstrates a useful connection between minimal sufficiency and the modal decomposition of CDK. 

\begin{proposition}[{\!\!\!\!\!\cite[Proposition 2]{xu2024dependence}}]
  \label{prop:mss}
  The features $\fss(X), \gss(Y)$ as defined by \eqref{eq:md:def}--\eqref{eq:fss:gss} are minimal sufficient statistics of $X$ and $Y$.
\end{proposition}

\section{Main Results}
Throughout our development in the paper, we will assume that $S = s(X), T = t(Y)$ are sufficient statistics of given $X, Y$, i.e., $X-s(X)-t(Y)-Y$. We demonstrate several common examples of information measures $\theta(X, Y)$ that can be evaluated from $S$ and $T$, i.e., $\theta(X, Y) = \theta(S, T)$.

\subsection{Mutual Information and $f$-Information}
It is straightforward to verify that the mutual information $$\ds I(X; Y) \defeq \sum_{x \in \cX, y\in \cY} P_{X, Y}(x, y) \log \frac{P_{X, Y}(x, y)}{P_X(x)P_Y(y)}$$ satisfies $I(X; Y) = I(S; T)$. To see this, note that we have $S-X-Y-T$ since $S$ and $T$ are deterministic functions of $X$ and $Y$, respectively. Then, the data processing inequality [cf. \lemref{lem:chain}] implies that $I(S; T) \leq I(X; Y)$. Also, the sufficiency of $S$ and $T$ implies that $X-S-T-Y$ and $I(X; Y) \leq I(S; T)$. %

We can extend the above result to the general class of $f$-information, defined as
 \begin{align}
   I_f(X; Y) \defeq \sum_{\substack{x \in \cX\\ y\in \cY}} P_{X}(x)P_Y(y) \cdot  f\left(\frac{P_{X, Y}(x, y)}{P_{X}(x)P_{Y}(y)}\right),
   \label{eq:f:info}
 \end{align}
where  %
 $f$ is a convex function defined on $[0, \infty)$ with $f(1) = 0$.
In particular the mutual information $I(X; Y)$ corresponds to the specific $f(u) = u\cdot \log u$. Then we have the following characterization, as a straightforward corollary of \propref{prop:suff}. We omit the proof.
 \begin{corollary}
   \label{cor:f:info}
   For each $f$-information defined by \eqref{eq:f:info}, we have $I_{f}(X; Y) = I_{f}(S; T)$.
\end{corollary} %

\subsection{Common Information}
We consider the common information between a pair of random variables formulated by Wyner \cite{wyner1975common} and the common information introduced by G{\'a}cs and K{\"o}rner \cite{gacs1973common}.

\subsubsection{Wyner's Common Information}
Given $X, Y$, Wyner's common information \cite{wyner1975common} between $X$ and $Y$ is defined as
\begin{align}
  \cw(X, Y) \defeq \min_{P_{W|X, Y}\colon X-W-Y} I(W; X, Y).
  \label{eq:wyner:def}
\end{align}

Then, we have the following characterization. A proof is provided in \secref{proof:thm:wyner}.

\begin{theorem}
\label{thm:wyner}
  Suppose $S = s(X)$ and $T = t(Y)$ are sufficient. Then, $\cw(X, Y) = \cw(S, T)$, and the optimal $W$ in \eqref{eq:wyner:def} satisfies $W - (S, T) - (X, Y)$.
\end{theorem}

\subsubsection{G{\'a}cs--K{\"o}rner Common Information}
Different from Wyner's notion, G{\'a}cs--K{\"o}rner common information 
\cite{gacs1973common} is %
\begin{align}
  \cgk(X, Y) \defeq \max_{f, g\colon \prob{f(X) = g(Y)} = 1} H(f(X)),\label{eq:def:cgk}
\end{align}
where $f\colon \cX \to \cW$ and $g \colon \cY \to \cW$ map $X$ and $Y$ to some common alphabet $\cW$, and where $H(\cdot)$ denotes Shannon entropy. Detailed analyses and properties of $\cgk(X, Y)$ can also be found in, e.g.,  \cite{witsenhausen1975sequences} and \cite[Section 5.9]{HuangMWZ2024}.

Here, we establish that $\cgk(X, Y)$ can be computed from sufficient statistics. A proof is provided in \secref{proof:thm:cgk}.

\begin{theorem}
  \label{thm:cgk}
    We have $\cgk(X, Y) = \cgk(S, T) = H(\mcf{0}(X), \dots, \mcf{k}(X))$, %
where we have defined $k \defeq \max\{0 \leq i \leq K\colon \sigma_i = 1\}$, and where $\sigma_i$'s and $K$ are defined by the modal decomposition \eqref{eq:md:def}. 
\end{theorem}

\subsection{Information Bottleneck} 

Given $X, Y$, the information bottleneck problem \cite{tishby2000information} 
investigates the description of $X$ by a new variable $U$, such that
$U$ captures as much of the information about $Y$ as possible. This can be formalized as solving the maximal $I(U; Y)$ under the constraint $I(U; X) \leq R$ for some $R \geq 0$, where $U$ is specified by $P_{U|X}$. This is equivalent to computing
   \begin{align}
    \vartheta_{X, Y}(R) \defeq \max_{\substack{P_{U|X} \colon U-X-Y\\ \quad I(U; X) \leq R}} I(U; Y).
     \label{eq:vartheta:R}
  \end{align}
From the data processing inequality, we have $\vartheta_{X, Y}(R) \leq I(X; Y)$. The equality can be achieved, e.g., for any $S = s(X)$ satisfy $X-S-Y$ and $R = H(S)$. %

\begin{remark}
    The information bottleneck formulation is a specific example of rate-distortion theory by employing a KL divergence as the distortion function \cite{tishby2000information}. The related quantities also appear in other information-theoretic studies. For example, $\vartheta_{X, Y}(R)$ is the optimal error exponent of a distributed hypothesis testing problem (cf. \cite[Theorem 3]{ahlswede1986hypothesis}), where $R$ corresponds to a communication rate constraint. 
  \end{remark}

The Lagrangian $I(U; X) - \beta \cdot I(U; Y)$ 
was introduced in \cite{tishby2000information} to characterize the tradeoff between $I(U; Y)$ and $I(U; X)$, where $\beta > 0$ is the Lagrange multiplier. Specifically, let us denote its minimum value as
\begin{align}
  \label{eq:ibls:def}
\!\!\!\!  \ibls(X, Y; \beta) \defeq \min_{P_{U|X} \colon U-X-Y} \left[I(U; X) - \beta \cdot I(U; Y)\right].
\end{align}

Compared with previous examples of information measures, $\vartheta(\cdot)$ and $\ibls$ are generally not symmetric with respect to $X$ and $Y$, and the results are parameterized functions instead of single scalar-valued measures. However, $\vartheta(\cdot)$ and $\ibls$ admit the same separable computation property, as demonstrated below. A proof is provided in \secref{proof:thm:ib}.

\begin{theorem}
\label{thm:ib}
Given $X, Y$, suppose $S = s(X), T = t(Y)$ are their sufficient statistics. Then, for all $\beta > 0$, we have $\ibls(X, Y; \beta) = \ibls(S, T; \beta)$, and the optimal $U$ that achieves $\ibls(X, Y; \beta)$ satisfies $U-S-X-Y$. In addition, $\vartheta_{X, Y}(R) = \vartheta_{S, T}(R)$ for all $R \geq 0$.
\end{theorem}

\section{Discussions}%
We demonstrated the separable design for estimating information measure $\theta$, where $\theta$ satisfies employing $\theta(X, Y) = \theta(S, T)$ for sufficient statistics $S$ and $T$ of $X$ and $Y$. This separability enables modular designs and often more efficient implementations. A practical choice of $S$ and $T$ can be the $\fss(X)$ and $\gss(Y)$ defined by the modal decomposition [cf. \eqref{eq:md:def}]. The detailed design of learning $\fss(X)$ and $\gss(Y)$ from the $X, Y$ samples was discussed in \cite{xu2024neural}. More general separable designs for feature-centric information processing systems, including multivariate cases, were also studied in \cite{xu2024neural}.

 Our developments are also deeply related to 
 the statistical dependence. For example, the relation $\theta(X, Y) = \theta(S, T)$, i.e., the information measure $\theta$ is invariance to the choices of sufficient statics $S$ and $T$, also appeared as a key criterion in characterizing the statistical dependence \cite{xu2024dependence}. In addition, \cite{HuangMWZ2024} demonstrated that
the features $\fss(X), \gss(Y)$, referred to as the universal features, are optimal for a large class of inference problems. In particular, under assumptions of weak dependence or Gaussian dependence, \cite{HuangMWZ2024} developed analytical expressions of information measures in terms of $\fss(X), \gss(Y)$. %

\section{Proofs}

We first introduce several useful results. The first is the data processing inequality, a proof of which can be found in, e.g., \cite[Theorem 2.8.1]{cover2006}.

\begin{lemma}
  \label{lem:chain}
  For $X, Y, Z$ with the Markov relation $X-Y-Z$, $I(X; Z) = I(X; Y) - I(X; Y|Z) \leq I(X; Y)$.
\end{lemma}

\begin{lemma}
  \label{lem:construct}
Given $X, U$ taking values from $ \cX \times\cU$ with joint distribution $P_{X, U}$, define $V \defeq v(X)$ for a function $v$ on $\cX$, and define a new random variable $U'$ on $\cU$ specified by $P_{U'|X}(u|x) \defeq P_{U|V}(u|v(x))$ for all $u \in \cU, x \in \cX$. Then we have $U' - V - X$ and $P_{U', V} = P_{U, V}$.
\end{lemma}
\begin{IEEEproof}
    Since $P_{U'|X}(u|x)$ depends on $x$ only through $v(x)$, we obtain $U' - V - X$, i.e., $V$ is a sufficient statistic of $X$ to infer $U'$. In addition, from $P_{U'|V}(u|v(x)) = P_{U|V}(u|v(x))$, we obtain $P_{U'|V} = P_{U|V}$. This implies that $P_{U'} = P_U$ and thus $P_{U', V} = P_{U, V}$.
  \end{IEEEproof}

\begin{lemma}
\label{lem:long:chain}
    If $U, X, Y, Z$ satisfy $U- X - Y$, $U-(X, Y)-Z$, and $X-Z-Y$, then we have $U-X-Z-Y$.
  \end{lemma}
\begin{IEEEproof}
Omitted.
\end{IEEEproof}

\subsection{Proof of \propref{prop:suff}}
\label{proof:prop:suff}
 First, note that ``2'' $\implies$ ``1'' can be readily obtained by taking corresponding marginal distributions. It suffices to prove that ``1'' $\implies$ ``3'' $\implies$ ``2''.

To establish ``1'' $\implies$ ``3'', from $X-S-Y$ we have
  \begin{align*}
    \frac{P_{X, Y}(x, y)}{P_{X}(x)P_Y(y)} = \frac{P_{Y|X}(y|x)}{P_{Y}(y)} = \frac{P_{Y|S}(y|s(x))}{P_{Y}(y)},
  \end{align*}
  which depends on $x$ only through $s(x)$.
  Similarly, from $X-T-Y$ we obtain 
  $$ \frac{P_{X, Y}(x, y)}{P_{X}(x)P_Y(y)} = \frac{P_{X|T}(x|t(y))}{P_{X}(x)},$$ 
which depends on $y$ only through $t(y)$. Therefore, there exists a function $\tau \colon \cS \times \cT \to \mathbb{R}$, such that 
  \begin{align}
    \frac{P_{X, Y}(x, y)}{P_{X}(x)P_Y(y)} = \tau(s(x), t(y)).
    \label{eq:tau}
  \end{align}

  This implies that for any $s_0 \in \cS, t_0 \in \cT$, 
  \begin{align*}
    P_{S, T}(s_0, t_0) &= \sum_{x \colon s(x) = s_0}\sum_{y \colon t(y) = t_0 }
P_{X, Y}(x, y)\\
    &= P_S(s_0)P_T(t_0) \cdot \tau(s_0, t_0).
  \end{align*}
  From \eqref{eq:tau} we obtain ``3'', since
  \begin{align*}
    \frac{P_{X, Y}(x, y)}{P_{X}(x)P_Y(y)} = \tau(s(x), t(y)) =     \frac{P_{S, T}(s(x), t(y))}{P_{S}(s(x))P_T(t(y))}.
  \end{align*}

  To establish ``3'' $\implies$ ``2'', note that
  \begin{align*}
    &P_{X, Y, S, T}(x, y, s(x), t(y))\\
    &\qquad= P_{X, Y}(x, y)\\
    &\qquad= P_{X}(x)P_Y(y) \cdot \frac{P_{S, T}(s(x), t(y))}{P_S(s(x))P_T(t(y))}\\
    &\qquad= \frac{P_{X}(x)}{P_S(s(x))} \cdot \frac{P_Y(y)}{P_T(t(y))} \cdot P_{S, T}(s(x), t(y))\\
    &\qquad= P_{X|S}(x|s(x)) \cdot P_{S, T}(s(x), t(y)) \cdot  P_{Y|T}(y|t(y)),
  \end{align*}
  which gives $X-S-T-Y$.
\hfill\IEEEQEDhere%

\subsection{Proof of \thmref{thm:wyner}}
\label{proof:thm:wyner}

We first introduce a useful lemma.
\begin{lemma}
\label{lem:wyner}
  Given $X, Y$, and suppose $S = s(X)$ and $T = t(Y)$ are sufficient. Then, for each random variable $W$ with $X-W-Y$ specified by the alphabet $\cW$ and the conditional distribution $P_{W|X, Y}$, we can construct $W'$ specified by $P_{W'|X, Y}$ such that: $W'$ takes values from $\cW$,
  \begin{itemize}
  \item $(X, Y) - (S, T) - W'$,  $P_{W'|S, T} = P_{W|S, T}$;
  \item $X - S - W' - T - Y$;
  \end{itemize}
\begin{align}
  I(W'; X, Y) 
  &= I(W; S, T)\notag\\
  &= I(W; X, Y) - I(W; X, Y|S, T).
    \label{eq:iw:eq}
\end{align}
\end{lemma}

\begin{IEEEproof}[Proof of \lemref{lem:wyner}]
  We construct $W'$ on $\cW$ as
 $P_{W'|X, Y}(w|x, y) \defeq P_{W|S, T}(w|s(x), t(y))$, for all $x \in \cX, y \in \cY$. %
From \lemref{lem:construct}, we obtain  $P_{W, S, T} = P_{W', S, T}$ and
\begin{align}
  \label{eq:w:ss}
  W' - (S, T) - (X, Y).
\end{align}
  We then verify that $X-S-W'-T-Y$, i.e.,
  \begin{align*}
&P_{X, Y, S, T|W'}(x, y, s(x), t(y)|w) \\
&~= P_{X|S}(x|s(x))P_{S|W'}(s(x)|w)P_{T|W'}(t(y)|w)P_{Y|T}(y|t(y)).
  \end{align*}
  To see this, note that %
  \begin{align}
    &
    P_{X, Y, S, T|W'}(x, y, s(x), t(y)|w)\notag\\
    &\quad= P_{X, Y|W'}(x, y|w)\notag\\
    &\quad= \frac{P_{W'|X, Y}(w|x, y)P_{X, Y}(x, y)}{P_{W'}(w)}\\
    &\quad= \frac{P_{W|S, T}(w|s(x), t(y))P_{X, Y}(x, y)}{P_{W}(w)}\\
    &\quad= P_{S, T|W}(s(x), t(y)|w) \cdot \frac{P_{X, Y}(x, y)}{P_{S, T}(s(x), t(y))}.\label{eq:longm}
  \end{align}
To simplify \eqref{eq:longm}, note that due to $X-W-Y$, we have the Markov relation $S-W-T$ and thus
  \begin{align}
    \!\!\!\!    P_{S, T|W}(s(x), t(y)|w) = P_{S|W}(s(x)|w) \cdot P_{T|W}(t(y)|w).
    \label{eq:longm:sub1}
  \end{align}
In addition, from the sufficiency of $S$ and $T$, we have
 $X-S-T-Y$. Thus, from \propref{prop:suff}, we have
  \begin{align}
    \frac{P_{X, Y}(x, y)}{P_{S, T}(s(x), t(y))}
    &= \frac{P_{X}(x)P_{Y}(y)}{P_{S}(s(x))P_T(t(y))}\notag\\
    &= P_{X|S}(x|s(x)) \cdot P_{Y|T}(y|t(y)).\label{eq:longm:sub2}
  \end{align}

Combining  \eqref{eq:longm:sub1} and   \eqref{eq:longm:sub2}, we can rewrite 
\eqref{eq:longm} as
  \begin{align}
   & P_{X, Y, S, T|W'}(x, y, s(x), t(y)|w)\notag\\
  &\quad= P_{S|W}(s(x)|w) \cdot P_{T|W}(t(y)|w)\notag\\
&\quad\qquad \cdot P_{X|S}(x|s(x))\cdot P_{Y|T}(y|t(y))\\
    &\quad= \left[ P_{X|S}(x|s(x)) \cdot P_{S|W'}(s(x)|w)\right] \notag\\
&\quad\qquad\cdot \left[P_{T|W'}(t(y)|w) \cdot P_{Y|T}(y|t(y))\right],
  \end{align}
where the last equality follows from that due to $P_{S, T|W} = P_{S, T|W'}$, we have $P_{S|W} = P_{S|W'}, P_{T|W} = P_{T|W'}$.

  Finally, \eqref{eq:iw:eq} follows from
    \begin{align}
      I(W'; X, Y) &= I(W'; S, T) \label{eq:mi:0}\\
      &= I(W; S, T) \label{eq:mi:1}\\
      &= I(W; X, Y) - I(W; X, Y|S, T)\label{eq:mi:4},
    \end{align}
  where %
  \eqref{eq:mi:0} follows from \eqref{eq:w:ss},
 \eqref{eq:mi:1} follows from the fact that $P_{W', S, T} = P_{W, S, T}$, and \eqref{eq:mi:4} follows from the Markov relation $W-(X, Y)-(S, T)$ and \lemref{lem:chain}.
\end{IEEEproof}

Proceeding to our proof of \thmref{thm:wyner}, from \lemref{lem:wyner}, for each $W$ with $X-W-Y$, we can construct $W'$ specified by $P_{W'|S, T}$ with $W'-(S, T)-(X, Y)$ and $X-S-W'-T-Y$ (which implies $X-W'-Y$), such that $I(W'; X, Y) = I(W; S, T) = I(W; X, Y) - I(W; X, Y|S, T)$. Therefore, the optimal $W$ satisfies $I(W; X, Y|S, T) = 0$, i.e., $W-(X, Y)-(S, T)$, since otherwise $W'$ is strictly better than $W$. Thus,
  \begin{align}
    \cw(X, Y) 
    &= \min_{P_{W|X, Y}\colon X-W-Y} I(W; X, Y)\notag\\
    &= \min_{\substack{P_{W|S, T}\colon W-(S, T)-(X, Y)\\\qquad\quad X-S-W-T-Y}} I(W; S, T).
    \label{eq:cw:xy}
  \end{align}

  In addition, for each $P_{W|S, T}$ such that $S-W-T$, we can construct $\Wb$ taking values from $\cW$, such that $P_{\Wb|X, Y}(w|x, y) = P_{W|S, T}(w|s(x), t(y))$. Then, we have $\Wb - (S, T) -(X, Y)$ and $P_{\Wb, S, T} = P_{W, S, T}$. This implies that $S-\Wb-T$ and $I(W; S, T) = I(\Wb; S, T)$.  Note that we also have $X-S-\Wb-T-Y$ since
  \begin{align*}
    &P_{X, Y, S, T|\Wb} (x, y, s(x), t(y)|w)\\
    &~= P_{X, Y| S, T,\Wb} (x, y| s(x), t(y), w) P_{S, T|\Wb}(s(x), t(y)|w)\\
    &~= P_{X, Y| S, T} (x, y| s(x), t(y)) P_{S, T|\Wb}(s(x), t(y)|w)\\
    &~= P_{X|S}(x|s(x)) P_{Y|T} (y|t(y)) P_{S|\Wb}(s(x)|w) P_{T|\Wb}(t(y)|w),
  \end{align*}
where the second equality follows from $\Wb - (S, T) -(X, Y)$, and where the last equality follows from $X-S-T-Y$ and $S-\Wb-T$. Therefore, we obtain
\begin{align}
  \cw(S, T) &= \min_{P_{W|S, T}\colon S-W-T} I(W; S, T)\notag\\
            &= \min_{\substack{P_{\Wb|S, T}\colon \Wb-(S, T)-(X, Y)\\\qquad\quad S-\Wb-T}} I(\Wb; S, T)\notag\\
            &= \min_{\substack{P_{\Wb|S, T}\colon \Wb-(S, T)-(X, Y)\\\qquad\quad X-S-\Wb-T-Y}} I(\Wb; S, T).
  \label{eq:cw:st}
\end{align}
Combining \eqref{eq:cw:xy} and \eqref{eq:cw:st}, we have $\cw(X, Y) = \cw(S, T)$. \hfill \IEEEQEDhere

\subsection{Proof of \thmref{thm:cgk}}
\label{proof:thm:cgk}
    From the modal decomposition \eqref{eq:md:def}, we define $\fh$ and $\gh$ as $\fh(x) \defeq (\mcf{0}(x), \dots, \mcf{k}(x))^\T, \gh(y) \defeq (\mcg{0}(y), \dots, \mcg{k}(y))^\T$. We first demonstrate that, for $d$-dimensional $f = (f_1, \dots, f_d)^\T$ and $g = (g_1, \dots, g_d)^\T$, $\prob{f(X) = g(Y)} = 1$ if and only if $f(x) = A \fh(x)$ and $g(y) = A \gh(y)$ for some  $A \in \mathbb{R}^{d \times (k+1)}$. Note that $\prob{f(X) = g(Y)} = 1$ is equivalent to $\prob{f_i(X) = g_i(Y)} = 1$ for all $i = 1, \dots, d$. Therefore, it suffices to consider the case $d = 1$.

To begin, we can uniquely decompose $f$ and $g$ as
\begin{align*}
  f = \sum_{i = 0}^{K} \alpha_i\cdot \mcf{i} + r_f,\quad
  g = \sum_{i = 0}^{K} \beta_i\cdot \mcg{i} + r_g,
\end{align*}
such that $\E{r_f(X)\mcf{i}(X)} = \E{r_g(Y)\mcg{i}(Y)} = 0$ for all $i = 0, \dots, K$. Then, it suffices to prove that
\begin{subequations}
  \label{eq:conds}
  \begin{gather}
      \alpha_i = \beta_i \quad \text{ for $i = 0, \dots, k$}, \\
      \alpha_i = \beta_i = 0 \quad \text{ for $i = k+1, \dots, K$}, \\
      r_f = 0, \quad r_g = 0.
\end{gather}
\end{subequations}
To this end, note that from the orthogonality between functions, 
\begin{align}
\!\!\!\!  &\E{(f(X) - g(Y))^2}\notag\\
  &\quad= \E{\left(\sum_{i = 0}^{K} \left(\alpha_i \mcf{i}(Y) - \beta_i \mcg{i}(Y)\right) + r_f(X) - r_g(Y)\right)^2}\notag\\
  &\quad=\sum_{i = 0}^{K} \E{\left(\alpha_i \cdot \mcf{i}(Y) - \beta_i \cdot \mcg{i}(Y)\right)^2}\notag\\
  &\qquad\qquad+ \E{r^2_f(X)} + \E{r^2_g(Y)}\notag\\
  &\quad=\sum_{i = 0}^{K} \left(\alpha_i^2 - 2\sigma_i \cdot \alpha_i\beta_i + \beta_i^2\right) + \E{r^2_f(X)} + \E{r^2_g(Y)}\notag\\
  &\quad=\sum_{i = 0}^{k} (\alpha_i - \beta_i)^2
 + \sum_{i = k+1}^{K} (1 - \sigma_i)(\alpha_i^2 + \beta_i^2)
\notag\\
 &\qquad + \sum_{i = k+1}^{K} \sigma_i(\alpha_i - \beta_i)^2+ \E{r^2_f(X)} + \E{r^2_g(Y)}.
\label{eq:E:dcmp}
\end{align}
Since $\prob{f(X) = g(Y)} = 1$ implies $\E{(f(X) - g(Y))^2} = 0$, from \eqref{eq:E:dcmp} we obtain the conditions \eqref{eq:conds}.

As a result, if $\prob{f(X) = g(Y)} = 1$, then
$H(f(X)) = H(A \fh(X)) \leq H(\fh(X)) = H(\mcf{0}(X), \dots, \mcf{k}(X))$, where the equality holds if $f = \fh$. Therefore, we obtain
\begin{align}
  \cgk(X, Y)
  &= \max_{f, g\colon \prob{f(X) = g(Y)} = 1} H(f(X))\notag\\
  &= H(\mcf{0}(X), \dots, \mcf{k}(X)).
  \label{eq:cgk:res}
\end{align}
From \propref{prop:suff}, we have
\begin{align*}
 \cdk_{S; T}(s(x), t(y)) = \sum_{i = 1}^K \sigma_i \cdot \mcf{i}(x) \cdot \mcg{i}(y),
\end{align*}
which is the modal decomposition of $\cdk_{S, T}$. Moreover, from the minimal sufficiency of $\fss(X), \gss(Y)$ [cf. \propref{prop:mss}], $\mcf{i}(x)$ is a function of $s(x)$, and $\mcg{i}(y)$ is a function of $t(y)$, for each $i = 0, \dots, K$. As a result, it follows from \eqref{eq:cgk:res} that $\cgk(S, T) = H(\mcf{0}(X), \dots, \mcf{k}(X)) = \cgk(X, Y)$.
\hfill \IEEEQEDhere

\subsection{Proof of \thmref{thm:ib}}
\label{proof:thm:ib}
We first introduce a useful lemma.
\begin{lemma}
\label{lem:u-s-y}
  Given $X, Y$, and suppose $S = s(X)$ satisfies $X-s(X)-Y$. Then, for each random variable $U$ with $U-X-Y$ specified by the alphabet $\cU$ and the conditional distribution $P_{U|X}$, we can construct $U'$ such that takes values from $\cU$ and satisfies $U'-S-X-Y$, $P_{U', S} = P_{U, S}$, $P_{U', Y} = P_{U, Y}$, and
  \begin{align*}
    I(U'; X) = I(U; X) - I(U; X|S),\quad I(U'; Y) = I(U; Y).
  \end{align*}
\end{lemma}
\begin{IEEEproof}[Proof of \lemref{lem:u-s-y}]
  We construct $U'$ on $\cU$ such that $P_{U'|X, Y}(u|x, y) = P_{U|S}(u|s(x))$. Then, it follows from \lemref{lem:construct} that  $U' - S - (X, Y)$ and $P_{U', S} = P_{U, S}$. Since we also have $S - X - Y$, we obtain 
$P_{U', S, X, Y} = P_{U', S} P_{X, Y|S} = P_{U', S} P_{X|S} P_{Y|X}$, i.e., $U'-S-X-Y$.

  Moreover, from $S-X-U$ and \lemref{lem:chain}, 
we have
  \begin{align}
    I(U; X)%
      = I(U; X|S) + I(U; S).
    \label{eq:i:exp1}
  \end{align}
  Similarly, we have 
  \begin{align}
    I(U'; X) = I(U'; X|S) + I(U'; S) = I(U'; S),
    \label{eq:i:exp2}
  \end{align}
  where the last equality follows from $U' - S - X$.

  From \eqref{eq:i:exp1}, \eqref{eq:i:exp2} and $P_{U,S} = P_{U',S}$, we obtain $I(U'; X) = I(U'; S) = I(U; S) = I(U; X) - I(U; X|S)$.

It remains only to establish $P_{U', Y} = P_{U, Y}$. To this end, note that from $U- X - Y$, $U-(X, Y)-S$, and $X-S-Y$, from \lemref{lem:long:chain} 
we obtain the Markov relation $U-X-S-Y$. Therefore, we have $U-S-Y$ and
  \begin{align*}
    P_{U, Y}(u, y) 
    =\sum_{s \in \cS}P_{Y|S}(y|s)P_{U, S}(u, s).
  \end{align*}   %
  In addition, due to $U'-S-X-Y$, we have $U'-S-Y$ and $$ P_{U', Y}(u, y) =\sum_{s \in \cS}P_{Y|S}(y|s)P_{U', S}(u, s).$$
 From $P_{U', S} = P_{U, S}$, we have $P_{U', Y} = P_{U, Y}$ and $I(U'; Y) = I(U; Y)$.
\end{IEEEproof}

Proceeding to the proof of \thmref{thm:ib}, we first show that
  \begin{align}
    \ibls(X, Y; \beta) = \ibls(S, Y; \beta) = \ibls(S, T; \beta).
    \label{eq:two:step}
  \end{align}

    To establish the first equality, note that the optimal $U$ that achieves $\ibls(X, Y; \beta)$ in \eqref{eq:ibls:def} satisfies $U-S-X$. Otherwise, $I(U; X|S) > 0$ and from \lemref{lem:u-s-y}, we can construct $U'$ on the same alphabet $\cU$ with $U'-S-X-Y$ and
 \begin{subequations}
   \label{eq:usxy}
   \begin{gather}
     I(U'; X) = I(U; X) - I(U; X|S) < I(U; X),\\
     I(U'; Y) = I(U; Y). 
   \end{gather}
 \end{subequations}
This implies that
  $I(U'; X) - \beta I(U; Y) < I(U; X) - \beta I(U; Y)$,
which contradicts the optimality of $U$.

Note that since we also have $U-X-Y$ and $X-(U, Y)-S$, from \lemref{lem:long:chain}, $U-S-X$ implies $U-S-X-Y$. Therefore, 
\begin{align*}
  \ibls(X, Y; \beta) %
   &= \min_{P_{U|X} \colon U-X-Y} \left[I(U; X) - \beta \cdot I(U; Y)\right]\\
  &= \min_{P_{U|S} \colon U-S-X-Y} \left[I(U; S) - \beta \cdot I(U; Y)\right]\\
  &= \min_{P_{U|S} \colon U-S-Y} \left[I(U; S) - \beta \cdot I(U; Y)\right]\\
  &= \ibls(S, Y; \beta),%
\end{align*}
where the second equality follows from that the optimal $U$ satisfies $U-S-X-Y$, and thus (cf. \lemref{lem:chain}) $I(U; X) = I(U; S) + I(U; X|S) = I(U; S)$. To obtain the third equality, note that for either $U-S-X-Y$ or $U-S-Y$, $I(U; S)$ and $I(U; Y)$ are solely determined by $P_{U|S}$. Finally, the first and the last equalities follow from the definition \eqref{eq:ibls:def}. 

To establish the second equality of \eqref{eq:two:step}, we note that $U- S- Y$ implies $U-S-T-Y$ and $I(U; Y) = I(U; T)$. Indeed, from $U- S- Y$, $U-(S, Y)-T$, and $S-T-Y$ (due to the statistical sufficiency of $S, T$), we obtain $U-S-T-Y$ by applying \lemref{lem:long:chain}. Therefore, we have $U-T-Y$ and $U-Y-T$, which implies that $I(U; Y) = I(U; T)$. Therefore, 
\begin{align*}
   \ibls(S, Y; \beta)
  &= \min_{P_{U|S} \colon U-S-Y} \left[I(U; S) - \beta \cdot I(U; Y)\right]\\
  &= \min_{P_{U|S} \colon U-S-Y-T} \left[I(U; S) - \beta \cdot I(U; T)\right]\\
  &= \min_{P_{U|S} \colon U-S-T} \left[I(U; S) - \beta \cdot I(U; T)\right],
\end{align*}
where the last equality follows from the fact that $I(U; S)$ and $I(U; T)$ depend only on $P_{U|S}$.

Similarly, we obtain
\begin{align}
  \vartheta_{X, Y}(R) &\defeq \max_{\substack{P_{U|X} \colon U-X-Y\\ \quad I(U; X) \leq R}} I(U; Y)\\
  &= \max_{\substack{P_{U|S} \colon U-S-X-Y\\ \quad I(U; S) \leq R}} I(U; Y)\label{eq:theta:1}\\
  &= \max_{\substack{P_{U|S} \colon U-S-Y\\ \quad I(U; S) \leq R}} I(U; Y)\label{eq:theta:2}\\
  &= \max_{\substack{P_{U|S} \colon U-S-Y-T\\ \quad I(U; S) \leq R}} I(U; T)\label{eq:theta:3}\\
  &= \max_{\substack{P_{U|S} \colon U-S-T\\ \quad I(U; S) \leq R}} I(U; T) =  \vartheta_{S, T}(R)\label{eq:theta:4}
\end{align}
where \eqref{eq:theta:1} follows from that it is without loss of generality to restrict $U$ to $U-S-X-Y$ [cf. \eqref{eq:usxy}], which satisfies $I(U; X) = I(U; S)$. \eqref{eq:theta:3} follows from that $U-S-Y$ implies $I(U; Y) = I(U; T)$. To obtain \eqref{eq:theta:2} and the first equality of  \eqref{eq:theta:4} we have dropped variables in the middle of Markov chains, as the objective functions ($I(U; Y)$ or $I(U; T)$) are uniquely determined by $P_{U|S}$. 
\hfill \IEEEQEDhere
\bibliographystyle{IEEEtran}
\bibliography{ref}

\begin{thebibliography}{10}
\providecommand{\url}[1]{#1}
\csname url@samestyle\endcsname
\providecommand{\newblock}{\relax}
\providecommand{\bibinfo}[2]{#2}
\providecommand{\BIBentrySTDinterwordspacing}{\spaceskip=0pt\relax}
\providecommand{\BIBentryALTinterwordstretchfactor}{4}
\providecommand{\BIBentryALTinterwordspacing}{\spaceskip=\fontdimen2\font plus
\BIBentryALTinterwordstretchfactor\fontdimen3\font minus
  \fontdimen4\font\relax}
\providecommand{\BIBforeignlanguage}[2]{{%
\expandafter\ifx\csname l@#1\endcsname\relax
\typeout{** WARNING: IEEEtran.bst: No hyphenation pattern has been}%
\typeout{** loaded for the language `#1'. Using the pattern for}%
\typeout{** the default language instead.}%
\else
\language=\csname l@#1\endcsname
\fi
#2}}
\providecommand{\BIBdecl}{\relax}
\BIBdecl

\bibitem{shannon1951prediction}
C.~E. Shannon, ``Prediction and entropy of printed english,'' \emph{Bell system
  technical journal}, vol.~30, no.~1, pp. 50--64, 1951.

\bibitem{belghazi2018mutual}
M.~I. Belghazi, A.~Baratin, S.~Rajeshwar, S.~Ozair, Y.~Bengio, A.~Courville,
  and D.~Hjelm, ``Mutual information neural estimation,'' in
  \emph{International conference on machine learning}.\hskip 1em plus 0.5em
  minus 0.4em\relax PMLR, 2018, pp. 531--540.

\bibitem{gowri2024approximating}
\BIBentryALTinterwordspacing
G.~Gowri, X.~Lun, A.~M. Klein, and P.~Yin, ``Approximating mutual information
  of high-dimensional variables using learned representations,'' in \emph{The
  Thirty-eighth Annual Conference on Neural Information Processing Systems},
  2024. [Online]. Available: \url{https://openreview.net/forum?id=HN05DQxyLl}
\BIBentrySTDinterwordspacing

\bibitem{xu2024dependence}
X.~Xu and L.~Zheng, ``Dependence induced representations,'' in \emph{2024 60th
  Annual Allerton Conference on Communication, Control, and Computing}.\hskip
  1em plus 0.5em minus 0.4em\relax IEEE, 2024, pp. 1--8.

\bibitem{wyner1975common}
A.~Wyner, ``The common information of two dependent random variables,''
  \emph{IEEE Transactions on Information Theory}, vol.~21, no.~2, pp. 163--179,
  1975.

\bibitem{gacs1973common}
P.~G{\'a}cs, , and J.~K{\"o}rner, ``Common information is far less than mutual
  information.'' \emph{Problems of Control and Information Theory}, vol.~2, pp.
  149--162, 1973.

\bibitem{tishby2000information}
N.~Tishby, F.~C. Pereira, and W.~Bialek, ``The information bottleneck method,''
  \emph{arXiv preprint physics/0004057}, 2000.

\bibitem{HuangMWZ2024}
\BIBentryALTinterwordspacing
S.-L. Huang, A.~Makur, G.~W. Wornell, and L.~Zheng, ``Universal features for
  high-dimensional learning and inference,'' \emph{Foundations and
  Trends{\textregistered} in Communications and Information Theory}, vol.~21,
  no. 1-2, pp. 1--299, 2024. [Online]. Available:
  \url{http://dx.doi.org/10.1561/0100000107}
\BIBentrySTDinterwordspacing

\bibitem{xu2024neural}
X.~Xu and L.~Zheng, ``Neural feature learning in function space,''
  \emph{Journal of Machine Learning Research}, vol.~25, no. 142, pp. 1--76,
  2024.

\bibitem{witsenhausen1975sequences}
H.~S. Witsenhausen, ``On sequences of pairs of dependent random variables,''
  \emph{SIAM Journal on Applied Mathematics}, vol.~28, no.~1, pp. 100--113,
  1975.

\bibitem{ahlswede1986hypothesis}
R.~Ahlswede and I.~Csisz{\'a}r, ``Hypothesis testing with communication
  constraints,'' \emph{IEEE transactions on information theory}, vol.~32,
  no.~4, pp. 533--542, 1986.

\bibitem{cover2006}
\BIBentryALTinterwordspacing
T.~M. Cover and J.~A. Thomas, \emph{Elements of information theory {(2.}
  ed.)}.\hskip 1em plus 0.5em minus 0.4em\relax Wiley, 2006. [Online].
  Available: \url{http://www.elementsofinformationtheory.com/}
\BIBentrySTDinterwordspacing

\end{thebibliography}

\end{document}